\newcommand{\roughly}[1]%
        {\mathrel{\raise.4ex\hbox{$#1$\kern-.75em\lower1ex\hbox{$\sim$}}}} 
\newcommand\bra[1]{\langle #1 |} 
\newcommand\ket[1]{|{#1}\rangle} 
\newcommand\ms{$\overline{\rm MS}$\ }

\newcommand\CO{{\cal O}} 
 
\newcommand\beq{\begin{eqnarray}} 
\newcommand\eeq{\end{eqnarray}}

\def\Dsl{\,\raise.15ex \hbox{/}\mkern-12.8mu D}

\newcommand\MeV{{\rm MeV\,}} 
\newcommand\GeV{{\rm GeV\,}}

\newcommand{\be}{\begin{equation}}
\newcommand{\ee}{\end{equation}}
\newcommand{\ber}{\begin{eqnarray}}
\newcommand{\eer}{\end{eqnarray}}

\documentstyle[preprint,epsfig,epsf,aps,floats,psfrag]{revtex}

\begin{document}
\tighten
\preprint{\vbox{
\hbox{INT-PUB 03-23}}}
\bigskip
\title{Heavy Quark $\bar{q}q$ Matrix Elements in the 
Nucleon from Perturbative QCD}
\author {Andrei Kryjevski \footnote{abk4@u.washington.edu}}
\address{Dept. of Physics and Institute for Nuclear Theory, 
University of Washington, Seattle, WA 98195}
\date{\today}
\maketitle

\begin{abstract}
The scalar heavy quark content of the nucleon, ${\cal M}_q = \bra{N} m_q \bar q q \ket{N},$ is relevant for computing
the interaction of dark matter candidates with ordinary matter, while ${\cal M}_s$ is important for predicting the properties 
of dense matter.
We compute ${\cal M}_q$ in perturbative QCD to ${\cal{O}}{(\alpha^3)}.$
As one goes from ${\cal M}_t$ to ${\cal M}_c$ the leading order contribution decreases as the number of light quarks is dropping,
while the radiative corrections grow and are all positive. The leading source of uncertainty in the calculation is due to 
the poorly known value of  ${\cal M}_s.$ A related calculation suggests that a large value for ${\cal M}_s$ may be reasonable.

\end{abstract}
 
   
\section {Introduction and Results}

Knowledge of matrix elements ${\cal M}_q \equiv \bra{N} m_q \bar q q \ket{N}$
provides us with some insight into the flavor properties of the 
nucleon wave function. 
Since scalars typically
couple to $m_q \bar{q} q,$ values of the quark mass operator matrix elements are also relevant for dark matter
searches, as well as for experiments searching for new weak forces. 
The value of ${\cal M}_s$
is important for kaon condensation in dense nuclear matter \cite{kaplan_nelson}, and its value has been the subject of a debate
\cite{gasser_leutw,jenkins_manohar}.

Heavy flavor ${\cal M}_q$'s have been calculated in perturbative QCD to the leading order in $\alpha$\cite{shif_vainst_zakharov}.
Using a technique similar to the ones developed in \cite{shif_vainst_zakharov,voloshin,chivukula} and making use of the four loop
beta function, anomalous mass dimension, and three loop heavy flavor threshold 
matching coefficients available in the literature \cite{chetyrkin}, in Section 2 we extend the calculation of 
heavy flavor ${\cal M}_q$'s to $\CO(\alpha^{3})$.
We find 
\pagebreak
\ber
{{\cal M}_c}&=&{\frac{2}{27}} \bigl(1 + 0.108 + 0.025 - 0.002 \bigr) M (1 - x_{uds}) = 78.61 (1 - x_{uds}) \,\MeV,\nonumber \\
{{\cal M}_b}&=&{\frac{2}{25}} \bigl(1 + 0.071 + 0.009 - 0.002 \bigr) M (1 - x_{uds}) = 80.92 (1 - x_{uds}) \,\MeV,\nonumber \\
{{\cal M}_t}&=&{\frac{2}{23}} \bigl(1 + 0.0402 + 0.0009 - 0.0007 \bigr) M (1 - x_{uds}) = 84.89 (1 - x_{uds}) \,\MeV, 
\label{numres_dlamdmq} 
\eer
where $M$ is the nucleon mass (we use $M = 938.3$ \MeV), and 
\ber
x_{uds} = \frac{\Sigma_{{\pi}N} +\Delta_s M }{M},
\label{xuds}
\eer
with $\Delta_s M$ being the shift in the nucleon mass which would result from setting $m_s=0,$
\ber
\Delta_s M = \int_0^{m_s} dy \frac{\partial M}{\partial m_s}\Biggl|_{m_s = y} = \int_0^{m_s} dy \,\bra{N} \bar s s \ket{N}|_{m_s = y}
\label{deltaMs}
\eer
and 
\ber
\Sigma_{{\pi}N}\equiv {{\left(m_u + m_d\right)}\over{2}}\bra{N} {\bar{u}{u} + \bar{d}{d}}\ket{N}\simeq 45\, \MeV.
\label{sigmapiN}
\eer
From (\ref{numres_dlamdmq}) we see that there are two competing effects as one goes from
the top to the charm quark matrix element: since the number of light quarks is dropping, the leading order contribution decreases, while 
the radiative corrections grow and are all positive. We also note that the greatest source of uncertainty in the above results is the poor 
knowledge of the strange matrix element entering $\Delta_s M.$

While the same technique is not applicable to ${\cal M}_s$ directly, since $m_s\sim \Lambda,$ the scale where QCD becomes strong, in 
Section 3 we calculate the strange quark matrix element in the QCD with a hypothetical heavy strange quark and 
massless $u$ and $d$ quarks and find indications
favoring a large value of ${\cal M}_s$
for the physical strange quark mass.

\section{The Six Flavor Calculation} 

We consider six flavor QCD ($n_f=6$) with three heavy quarks 
($c,\,b,\,t$) and three light quarks ($u,\,d,\,s$). The running heavy quark masses are
 defined to be $\bar m_h(\mu)$ in the \ms scheme, and we define the
 heavy quark mass 
 $m_h$ to be the ``scale invariant mass'', namely the solution to the
 equation $\bar m_h(m_h)=m_h$. 

 For $\mu>m_h$ the theory is described by the QCD Lagrangian with 
 $n_f$ ``active'' flavors; at
 the scale $\mu=m_h$ we integrate out the heavy quark $h$, and until the next heavy flavor threshold
 the effective theory consists of $n_f-1$ flavor QCD, with a shifted gauge
 coupling, and a tower of nonrenormalizable interactions suppressed by
 powers of $m_h^2.$ The $\CO\left(1/{m_h^2}\right)$ contribution, for example, comes from the operator 
${(D_{\mu}G^{\mu\nu})^2}/{m^2_h} $ in the effective theory \cite{shif_vainst_zakharov}.
 The value of the coupling and of the coefficients
 of the nonrenormalizable interactions are fixed in perturbation
 theory by matching $S$-matrix elements in the full and effective
 theories. The effective theory is asymptotically free, and gets strong at the
 scale $\Lambda$. 

The mass of the nucleon $M$ is given by an expansion in $m_h^{-2}$ and $m_l$ of the form
\ber
M = \Lambda \times \left[c_0 + \sum_{n=1}^{\infty}
{\sum_{h} c_{n\,h}{\left(\frac{\Lambda^2}{m_h^2}\right)}^n} +
\sum_{m=1}^{\infty}\sum_{n=0}^{\infty} {\sum_{h,l} c_{n\,m,h\,l} \left({\frac{m_l}{\Lambda}}\right)^m 
{\left(\frac{\Lambda^2}{m_h^2}\right)}^n}\right] + \Delta_s M,
\label{mass_h}
\eer
where $h$ is a generic index for the heavy quarks, $l$ is a generic index for $u$ and $d$, and $\Delta_s M$ is defined in (\ref{deltaMs}).
The expansion coefficients 
depend on the details of the dynamics of the theory
and may depend logarithmically on the ratio
$\Lambda^2/m_h^2$ and/or $m_l/\Lambda$. 
We perform an expansion in $m_l/\Lambda$ since chiral perturbation theory works well for $u$ and $d$ quarks. The rationale for treating 
the strange quark differently than the light or heavy flavors is that we do not want to assume that either the heavy quark or chiral 
expansions converge well for the strange quark.

We define a fixed reference scale $\mu_0>m_t$ 
with $\alpha(\mu_0)\equiv \alpha_0.$ The scale $\Lambda$ is defined by keeping $\alpha(\Lambda) \equiv \alpha_{\Lambda}$ fixed; the 
value itself is unimportant to us, only its logarithmic derivative matters. Holding gauge couplings $\alpha_0$ and $\alpha_{\Lambda}$ 
fixed, we can invoke the Feynman-Hellman theorem to compute the $\bar q q$ matrix element in the nucleon 
by differentiating $M$ with respect to $\bar m_q(\mu_0)$:
\ber
\bar m_q(\mu_0) \frac{\partial M}{\partial \bar m_q(\mu_0)}\Biggl|_{\alpha_0,\,\alpha_{\Lambda}} &=&
\bar m_q(\mu_0) \frac{\partial}{\partial \bar m_q(\mu_0)} \bra{N}{H_{QCD}}\ket{N}\cr
 &=&
\bra{N} \bar m_{q}(\mu) \bar q q \ket{N}.
\label{fhh}
\eer
The equality above follows from the
Feynman-Hellman theorem since $m_q$ enters the QCD Hamiltonian in the
combination $m_q\bar q q$.
We have not specified a renormalization scale in the matrix elements as they are renormalization scale invariant.

\begin{figure}[t]
\centering{
\begin{psfrags}
\psfrag{y}{$\alpha^{-1}$}
\psfrag{x}{$\mathbf{\ln\,\mu}$}
\psfrag{lnmt}{$\ln \,m_t$}
\psfrag{a0}{$\alpha_0^{-1}$}
\psfrag{al}{$\alpha_{\Lambda}^{-1}$}
\psfrag{lnm0}{$\ln \,\mu_0$}
\psfrag{lnmb}{$\ln \,m_b$}
\psfrag{lnmc}{$\ln \,m_c$}
\psfrag{L}{$\ln \,\Lambda$}
\epsfig{figure=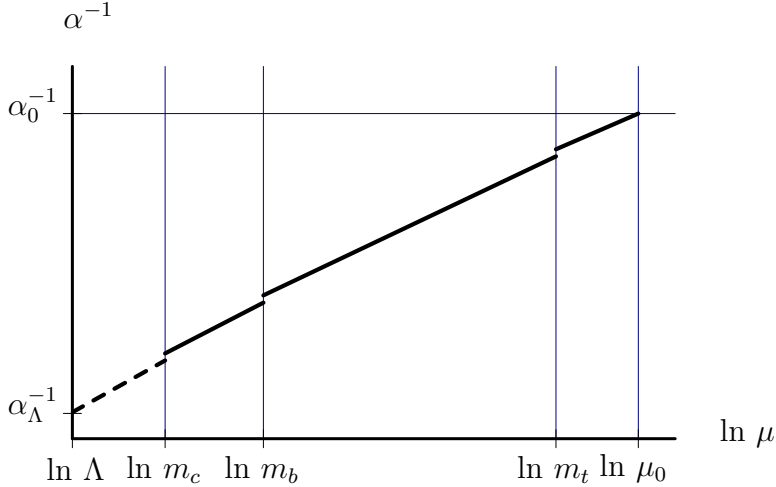,width=.65\textwidth}
\end{psfrags}
}
\caption{Shown here is a schematic plot of the QCD coupling $\alpha(\mu)^{-1}$ versus $\ln\,\mu,$ in the \ms scheme. The threshold 
corrections have been exaggerated for illustrative purposes. Keeping 
$\mu_0,\,\alpha_0$ and $\alpha_{\Lambda}$
fixed and varying $\bar m_{q}(\mu_0)$ one varies the value of $\Lambda$ as well the locations of (some of) 
the heavy flavor thresholds $\bar m_h(m_h)=m_h$.}
\label{fig1}
\end{figure}

Let us first consider the chiral limit $m_u = m_d = m_s = 0.$ Approximating (\ref{mass_h}) with
\ber
M = c_0 \Lambda + \CO\left(\frac{\Lambda^3}{m_h^2}\right)
\label{Mchiral}
\eer
we can express the heavy matrix element as
\ber
 {\cal M}_q 
&=& M \frac{\bar m_{q}(\mu_0)}{\Lambda} \frac{\partial \Lambda}{\partial
   \bar m_{q}(\mu_0)}\Biggl|_{\alpha_0,\,\alpha_{\Lambda}}\left[1 + \CO\left(\frac{\Lambda^3}{M m_h^2}\right)\right].
\label{master_h}
\eer

Thus to leading order in $\Lambda^3/M m_h^2$ the heavy quark matrix
element is determined by how $\Lambda$ varies when we vary the
``bare'' quark mass $\bar m_{q}(\mu_0)$, keeping  $\alpha_{\Lambda}$ and the ``bare'' gauge coupling $\alpha_0$
fixed (see Fig. 1).
This calculation requires
 knowledge of the anomalous mass dimension,
 the beta function for the running of the gauge coupling, and
 the matching functions that relate 
 the gauge coupling and the running quark masses in the full $n_f$ and the effective $(n_f-1)$-flavor gauge
 theories. 
Here we use the notation and results compiled in the paper by Chetyrkin, Kuhn, and 
Steinhauser\cite{chetyrkin} to which we refer reader for the formulas we use and the references to the original results.

The coupling constant $a(\mu)\equiv\alpha^{(n_f)}(\mu)/\pi$  in the theory with $n_f$ ``active'' flavors 
runs with the scale $\mu$ according to
the equation
\ber
\mu^2\frac{d a\ }{d\mu^2} &=& \beta^{(n_f)}(a),
\label{alpharun}
\eer
where the QCD beta function, $$\beta^{(n_f)}(a)= - \sum_{k \geq 0} \beta^{(n_f)}_k a^{k+2},$$ has been 
calculated to four loop order and the coefficients $\beta^{(n_f)}_k$ are known for $k = 0,1,2,3$. 
The current quark mass $m^{(n_f)}_q(\mu)$ of the theory with $n_f$ active flavors obeys
\ber 
\mu^2\frac{d\ }{d\mu^2} \bar m^{(n_f)}_q(\mu) &=& \bar m^{(n_f)}_q(\mu) \gamma_{m}^{(n_f)} (a), 
\label{mqrun}
\eer
where the anomalous mass dimension, $$\gamma_{m}^{(n_f)} (a)= - \sum_{k \geq 0} \gamma_{m,k}^{(n_f)} a^{k+1},$$ has been calculated to 
four loop order and the coefficients $\gamma^{(n_f)}_{m,k}$ are known for $k = 0,1,2,3.$ 
At a heavy quark threshold $\mu=m_h$ 
the couplings and the quark running masses in the ``full'' and the effective theories are related by the matching conditions
\ber
a^{(n_f-1)}(m_h)&=&a^{(n_f)}(m_h)\,\zeta^2_g(a^{(n_f)}(m_h)),\nonumber \\
\bar m_q^{(n_f-1)}(m_h) &=& \bar m_q^{(n_f)}(m_h)\,\zeta_m(a^{(n_f)}(m_h)).
\label{match_cond}
\eer
The functions $\zeta^2_g$ and $\zeta_m$ have been computed in perturbation theory
to three loop order.
Here 
$a^{(n_f)}(m_h), \bar m_{q}^{(n_f)}(m_h)$ and $a^{(n_f-1)}(m_h), \bar m_{q}^{(n_f-1)}(m_h)$ are the coupling and quark masses evaluated  
just above and below the flavor threshold, respectively. 

Equations (\ref{alpharun}) and (\ref{mqrun}) may be integrated to give
\ber
{\mathrm ln}[\frac{\mu_0^2}{\Lambda^2}] = {\int^{a(\mu_0)}_{a(\Lambda)}}
  \frac{d a}{\beta^{(n_f)}(a)},
\label{intalpharun}
\eer

\ber
{\mathrm \ln} \bigl[ \frac {\bar m_h(\mu_0)}{m_h} \bigr] = 
{\int^{a(\mu_0)}_{a(m_h)}}{d{a}{\frac{\gamma_m^{(n_f)}(a)}{\beta^{(n_f)}(a)}}}.
\label{intmqrun}
\eer

In order to obtain the desired quantity in (\ref{master_h}) we differentiate
(\ref{intalpharun}) with respect to 
$\ln \,\bar m_q(\mu_0)$, keeping $\mu_0$, $\alpha_0$, 
and $\alpha_{\Lambda}$ fixed and taking into account the flavor threshold discontinuities. 
The result is
\ber
\frac {\partial \, \ln \, \Lambda} {\partial \ln \,\bar m_q(\mu_0)} & = &
\sum_{h} 
\left(1 - {\frac {\beta^{(n_f)}(a_{{n_f}h})} {\beta^{(n_f-1)}(\zeta_g^2(a_{{n_f}h})a_{{n_f}h})}}
\frac{d \,\zeta_g^{2} (a_{{n_f}h})a_{{n_f}h}}{d a_{{n_f}h}} \right)
\frac{\partial \,\ln m_h}{\partial \,\ln \, \bar m_q(\mu_0)}
\label{dlamdm0h}
\eer
Here $a_{{n_f}h} \equiv a^{(n_f)}(m_h)$ and
the sum is over the heavy flavors lighter than the flavor $q$ and the $q$ itself.
To calculate ${{\partial \ln m_h}/{\partial\ln\,\bar m_q(\mu_0)}}$ we differentiate (\ref{intmqrun}) with respect 
to $\ln\,\bar m_q(\mu_0)$  keeping $\mu_0$, $\alpha_0$ 
fixed and taking into account the flavor threshold discontinuities in the running quark masses and coupling 
constant. 

Now let us take into account the light quark masses. Working in the isospin limit $m_u = m_d = m_l,$ we approximate (\ref{mass_h}) with
\ber
M = c_0 \Lambda + 2\,c_{1\,0\,h\,l}\, m_l + \Delta_s M + \CO\left(\frac{\Lambda^3}{m_h^2},\frac{m_l^2}{\Lambda}\right)
\label{Mml}
\eer
and the heavy matrix element expression is modified to
\ber
 {\cal M}_q 
&=& M \left(1- x_{u\,d\,s}\right)\frac{\bar m_{q}(\mu_0)}{\Lambda} \frac{\partial \Lambda}{\partial
\bar m_{q}(\mu_0)}\Biggl|_{\alpha_0,\alpha_{\Lambda}}\left[1 + \CO\left(\frac{\Lambda^3}{M m_h^2},{\frac{m_s^2}{M\,\Lambda}}\right)\right],
\label{master_h_ml}
\eer
where $x_{u\,d\,s}$ is defined in (\ref{xuds});
$\Sigma_{{\pi}N}\simeq 45\,$ \MeV is defined in (\ref{sigmapiN}) and is known
from pion-nucleon scattering, while 
the value of the strange matrix element, ${\cal M}_s \simeq 200 \,\MeV,$ that enters $\Delta_s M$ from (\ref{deltaMs}) is quite uncertain 
and varies by hundreds of \MeV depending on the method of calculation \cite{gasser_leutw,jenkins_manohar}.

We used the 
following values of parameters: $M_Z = 91.18$ \GeV, $\alpha(M_Z) = 0.117,$ 
$m_c = 1.2$ \GeV,
$m_b = 4.2$ \GeV and $m_t = 175$ \GeV \cite{pdg}. The 
numerical results obtained are shown in (\ref{numres_dlamdmq}).

\section{Calculation in the Three Flavor Toy Model}
The value of ${\cal{M}}_s$ has been calculated in baryon chiral perturbation theory.
The leading order calculation
predicts linear growth of the matrix element as a function of $m_s,$ however, at one loop order one 
gets sizable negative contributions from the meson loops proportional to $m_s^{3/2}$ \cite{gasser_leutw,jenkins_manohar}. This indicates 
that chiral perturbation theory may not be the right tool for the calculation. Calculations of ${\cal{M}}_s$ in the Skyrme model lead 
to the same conclusion \cite{kaplan_klebanov}. This is unfortunate since the value of ${\cal{M}}_s$ may have important consequences for 
the nature of dense matter \cite{kaplan_nelson}, and the fate of supernova remnants \cite{bethe_brown}. It could also be relevant for 
the properties of the strongly bound kaonic system $\mathrm{K}^{-}ppn$ observed recently \cite{iwasaki}. 

One may, however, gain qualitative insight into this problem using a rather different approach. We calculate the strange matrix element in 
three flavor QCD with a hypothetical heavy strange quark (such that $m_s\gg \Lambda$) 
and massless $u$ and $d$ quarks using the technique of the previous section.

In the effective $n_f=2$ theory the mass of the nucleon $M$ is given by an expansion in $m_s^{-2}$ of the form
\ber
M = \Lambda \times \sum_{n=0}^\infty  c_n
\left(\frac{\Lambda^2}{m_s^2}\right)^n\ ,
\label{mass}
\eer
where the coefficients $c_n$ will depend logarithmically on the ratio
$\Lambda/m_s$. 
Defining a reference scale $\mu_0>m_s$ with $\bar m_s(\mu_0)\equiv m_{s0}$ and gauge coupling $\alpha(\mu_0)\equiv \alpha_0,$
and using the same line of argument as in the previous section, 
we can express the strange matrix element as
\ber
{\cal M}_s 
= M
 \frac{m_{s0}}{\Lambda} \frac{\partial \Lambda}{\partial
   m_{s0}}\Biggl|_{\alpha_0,\,\alpha_{\Lambda}}
\left[1 +\CO\left(\frac{\Lambda^3}{m_s^2 M}\right)\right]\ .
\label{master}
\eer
Thus to leading order in $\Lambda^3/M m_s^2$ the strange matrix
element is determined by how $\Lambda$ varies when we vary the
``bare'' strange quark mass, keeping $\alpha_{\Lambda}$ and the ``bare'' gauge coupling $\alpha_0$
fixed.

\begin{figure}[t]
\centering{
\begin{psfrags}
\psfrag{y}{${{\cal M}_s}/M$}
\psfrag{X}{$m_s,$ \GeV}
\epsfig{figure=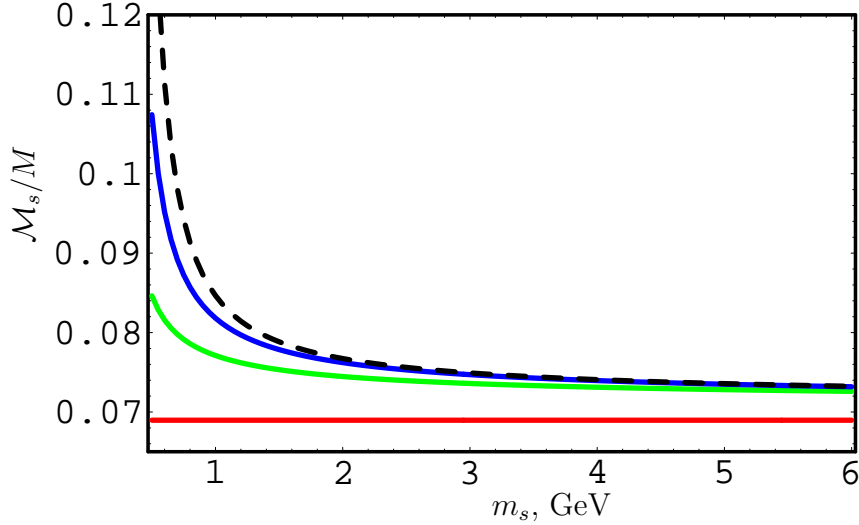,width=.70\textwidth}
\end{psfrags}
}
\caption{
Shown here is the perturbative result for ${\cal M}_s/M \equiv \bra{N} m_s \bar s s \ket{N} / M$ versus $m_s\,$ in \GeV calculated in the 
toy three flavor model with 
massless $u$ and $d$ quarks and a heavy $s$ quark.
At the leading order ${\cal M}_s/M = 2/29 \simeq 0.069$ (the red line); the green line corresponds to the NLO result, the blue line - NNLO 
and 
the black dashed line - NNNLO. Perturbation theory is clearly breaking down for $m_s\leq 1$ \GeV.}
\label{fig2}
\end{figure}

In order to obtain the desired quantity in (\ref{master}) we
differentiate both sides of (\ref{intalpharun}) and (\ref{intmqrun}) with respect to $\mathrm{ln}\,m_{s0}$, keeping $\mu_0$, 
$\alpha_0$, and $\alpha_{\Lambda}$ fixed and taking into account the strange quark threshold discontinuity. We get
\ber
 & &\frac{1}{M}\bra{N} m_s \bar s s \ket{N}=\frac{\partial\ln \,\Lambda}{\partial \ln \,m_{s0}}\Biggr|_{\alpha_0,\alpha_{\Lambda}}=
\cr&=&\frac{\partial\ln \,m_s}{\partial\ln \, m_{s0}}
\left[1-\frac{\beta^{(3)}(a)}{\beta^{(2)}(\zeta_g^2(a)\,a)} \frac{d}{d a}{(\zeta_g^2(a)\,a)}\right]=\cr
&=&
\left[\frac{1}{1-2\gamma^{(3)}_m(a)}\right]
\left[1-\frac{\beta^{(3)}(a)}{\beta^{(2)}(\zeta_g^2(a)\,a)} \frac{d}{d a}(\zeta_g^2(a)\,a)\right],
\label{dlamdms}
\eer
where $a\equiv{a^{(3)}(m_s)}$. 
The coupling ${a^{(3)}(m_s)}$ is calculated by running down from $\mu = M_Z$ to $\mu = m_s$ using (\ref{intalpharun}) with the real world 
initial condition 
$\alpha(M_Z) = 0.117.$

Fig. 2 shows ${\cal M}_s/M \equiv \bra{N} m_s \bar s s \ket{N} / M$ as a function of $m_s$ 
calculated through four loop order in perturbation theory. It indicates that QCD may favor a bigger value of the strange matrix element 
for the physical $m_s.$ A reasonable guess for ${\cal{M}}_s(m_s)$ in the nonperturbative regime consistent with what we know about 
${\cal{M}}_s$ for the values of $m_s,$ both larger and smaller than its physical value, is sketched in Fig. 3.

\begin{figure}[t]
\centering{
\begin{psfrags}
\psfrag{y}{${{\cal M}_s}$}
\psfrag{x}{$m_s$}
\psfrag{L}{$\Lambda$}
\psfrag{gev}{1 \GeV}
\epsfig{figure=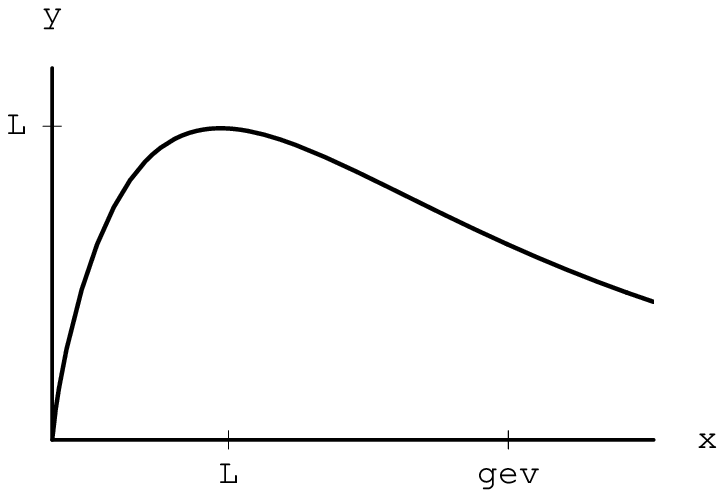,width=.60\textwidth}
\end{psfrags}
}
\caption{}
\label{fig3}
\end{figure}
\bigskip
\begin{center}
\large{ \textbf{Acknowledgments}}
\end{center}
I thank D. B. Kaplan for many helpful discussions along the way. 
The work is supported by the US Department of Energy grant
DE-FG03-00ER41132.

\newcommand{\IJMPA}[3]{{ Int.~J.~Mod.~Phys.} {\bf A#1}, (#2) #3}
\newcommand{\JPG}[3]{{ J.~Phys. G} {\bf {#1}}, (#2) #3}
\newcommand{\AP}[3]{{ Ann.~Phys. (NY)} {\bf {#1}}, (#2) #3}
\newcommand{\NPA}[3]{{ Nucl.~Phys.} {\bf A{#1}}, (#2) #3 }
\newcommand{\NPB}[3]{{ Nucl.~Phys.} {\bf B{#1}}, (#2)  #3 }
\newcommand{\PLB}[3]{{ Phys.~Lett.} {\bf B{#1}}, (#2) #3 }
\newcommand{\PRv}[3]{{ Phys.~Rev.} {\bf {#1}}, (#2) #3}
\newcommand{\PRC}[3]{{ Phys.~Rev. C} {\bf {#1}}, (#2) #3}
\newcommand{\PRD}[3]{{ Phys.~Rev. D} {\bf {#1}}, (#2) #3}
\newcommand{\PRL}[3]{{ Phys.~Rev.~Lett.} {\bf {#1}}, (#2) #3}
\newcommand{\PR}[3]{{ Phys.~Rep.} {\bf {#1}}, (#2) #3}
\newcommand{\ZPC}[3]{{ Z.~Phys. C} {\bf {#1}}, (#2) #3}
\newcommand{\ZPA}[3]{{ Z.~Phys. A} {\bf {#1}}, (#2) #3}
\newcommand{\JCP}[3]{{ J.~Comput.~Phys.} {\bf {#1}}, (#2) #3}
\newcommand{\HIP}[3]{{ Heavy Ion Physics} {\bf {#1}}, (#2) #3}
\newcommand{\RMP}[3]{{ Rev. Mod. Phys.} {\bf {#1}}, (#2) #3}
\newcommand{\APJ}[3]{{Astrophys. Jl.} {\bf {#1}}, (#2) #3}
\newcommand{\LNP}[3]{{Lect. Notes Phys.} {\bf {#1}}, (#2) #3}
\newcommand{\RNC}[4]{{Riv. Nuovo Cim.} {\bf {#1}N{#2}}, (#3) #4}
\newcommand{\JHP}[3]{{ JHEP} {\bf {#1}}, (#2) #3}
\newcommand{\CPC}[3]{{Comput. Phys. Commun} {\bf {#1}}, (#2) #3}


\begin{thebibliography}{99}

\bibitem{kaplan_nelson}
D.B.~Kaplan and A.E.~Nelson, \NPA{479}{1988}{273}

\bibitem{gasser_leutw}
J.~Gasser, Ann. \ Phys. {\bf 136},\,(1981) 62; J.~Gasser, H.~Leutwyler and M.E.~Sainio, \PLB{253}{1991}{252}

\bibitem{jenkins_manohar}
E.~Jenkins and A.V.~Manohar,\PLB{281}{1992}{336}

\bibitem{shif_vainst_zakharov}
M.A.~Shifman, A.I.~Vainstein and V.I.~Zakharov,\PLB{78}{1978}{443}

\bibitem{voloshin}
M.B.~Voloshin, Yad.\ Fiz. {\bf 45}, 190 (1987)

\bibitem{chivukula}
R.S.~Chivukula, A.~Cohen, H.~Georgi and A.~Manohar, \PLB{222}{1989}{258}

\bibitem{chetyrkin}
K.G. Chetyrkin, Johann H. Kuhn, and M. Steinhauser, \CPC{133}{2000}{43-65} 

\bibitem{pdg}
K.~Hagiwara et al., \PRD {66}{2002}{010001}

\bibitem{kaplan_klebanov}
D.~B.~Kaplan, and I.~R.~Klebanov, \NPB{335}{1990}{45}

\bibitem{bethe_brown}
G.~E.~Brown, and H.~Bethe,\, Astrophys.\, J.\, {\bf 423},\,(1994)\, 659

\bibitem{iwasaki}
M.~Iwasaki $\it{et.}$ $\it{al.}$, nucl-ex/0310018

\end{thebibliography}
\end{document}